\documentclass[a4paper]{article}

\usepackage[english]{babel}
\usepackage[utf8x]{inputenc}
\usepackage[T1]{fontenc}
\usepackage{gensymb} 
\usepackage{threeparttable}
\usepackage{tabularx}
\usepackage{makecell}
\usepackage{cite}
\usepackage{mathrsfs}
\usepackage{amssymb}
\usepackage{xcolor}
\usepackage{booktabs}
\usepackage{textcomp}
\usepackage{stfloats}
\usepackage{pifont}
\makeatletter
\newcommand{\ssymbol}[1]{^{\@fnsymbol{#1}}}
\makeatother
\newtheorem{theorem}{Theorem}{}

\usepackage{subcaption}
\usepackage{amsmath}
\usepackage{graphicx}
\usepackage[colorinlistoftodos]{todonotes}
\usepackage[colorlinks=true, allcolors=blue]{hyperref}

\usepackage{caption}
\usepackage[a4paper,top=2cm,bottom=2.5cm,left=2.5cm,right=2.5cm,marginparwidth=1.75cm]{geometry}

\usepackage{authblk}
\title{ Quantum Scheme for Private Set Intersection and Union Cardinality based on Quantum Homomorphic Encryption}
\author[1]{Chong-Qiang Ye}
\author[2]{Jian Li}
\author[3]{Tianyu Ye}
\author[1]{Xiaoyu Chen \thanks{Corresponding author: Xiaoyu Chen, Email: chenxiaoyu@hzcu.edu.cn}}

\affil[1]{ School of Information and Electrical Engineering, Hangzhou City University, Hangzhou 310015, China}
\affil[2]{School of Cyberspace Security, Beijing University of Posts and Telecommunications, Beijing 100876, China }
\affil[3]{College of Information \& Electronic Engineering, Zhejiang Gongshang University, Hangzhou 310018, China }
\begin{document}
\date{}

  \maketitle

\begin{abstract}
Private set intersection (PSI) and private set union (PSU) are the crucial primitives in secure multiparty computation protocols, which enable several participants to jointly compute the intersection and union of their private sets without revealing any additional information. Quantum homomorphic encryption (QHE) offers significant advantages in handling privacy-preserving computations. However, given the current limitations of quantum resources, developing efficient and feasible QHE-based protocols for PSI and PSU computations remains a critical challenge.
In this work, a novel quantum private set intersection and union cardinality protocol is proposed, accompanied by the corresponding quantum circuits. Based on quantum homomorphic encryption, the protocol allows the intersection and union cardinality of users' private sets to be computed on quantum-encrypted data with the assistance of a semi-honest third party. By operating on encrypted quantum states, it effectively mitigates the risk of original information leakage. Furthermore, the protocol requires only simple Pauli and CNOT operations, avoiding the use of complex quantum manipulations (e.g., $T$ gate and phase rotation gate). Compared to related protocols, this approach offers advantages in feasibility and privacy protection.
\\
\\
\textbf{Keywords:} Quantum communication, Secure multiparty computation, Homomorphic encryption, Privacy protection

\end{abstract}

\section{Introduction}

In today's digital era, data privacy and security have become paramount. Protecting user data against unauthorized access and breaches poses significant challenges in modern computing. As technological advancements continue, ensuring robust privacy protection mechanisms is increasingly crucial.

Private set intersection (PSI) and private set union (PSU) are essential primitives in secure multiparty computation protocols \cite{1,2}. PSI allows parties to jointly compute the common elements of their private sets without disclosing any additional information, while PSU enables them to compute the union of their sets. These primitives are critical for privacy-preserving data sharing, collaborative computation, and secure database operations in fields such as cybersecurity, healthcare, and finance\cite{3,4}. To further reduce the leakage of private information, private set intersection cardinality (PSI-CA) and private set union cardinality (PSU-CA) were developed\cite{5,6,7}. In these variants, parties compute only the size of the intersection or union, rather than revealing the actual elements.

%
However, with the rise of quantum computing \cite{8,9}, the security and efficiency of traditional PSI (PSU) protocols need to be re-evaluated. Quantum computers have enormous computing power and can efficiently solve problems that are currently intractable with traditional machines, such as factoring large integers and solving discrete logarithms \cite{10}, thereby breaking many of the classical encryption algorithms that these protocols rely on, such as RSA and ECC\cite{11}. To counter the threats posed by the powerful computational capabilities of quantum machines, researchers are exploring the integration of quantum encryption technologies into classical PSI (PSU) schemes.

Quantum private set intersection (QPSI) and quantum private set union (QPSU), are designed to withstand quantum attacks and have thus garnered significant attention and extensive research. In 2016, Shi et al. \cite{12} proposed a deception-sensitive QPSI protocol that addresses the intersection of private datasets among users through encoded quantum states, quantum operations, and von Neumann measurements. To calculate the cardinality of the private set intersection, Shi et al. \cite{13} designed a quantum private set intersection cardinality protocol using quantum Fourier transform and quantum counting algorithm. Then, Zhang et al.\cite{14} designed a quantum privacy set intersection and union cardinality solution for three-party scenarios by utilizing GHZ states. However, these protocols rely on multi-particle entangled states or complex quantum oracle operators, which are difficult to achieve under current quantum technology. In 2021, Liu \cite{15} proposed an improved QPSI protocol based on single photons. By using single photons as the information carrier, the protocol reduces both the complexity and cost of implementation. In 2023, Mohanty \cite{16} proposed a multi-party quantum PSI (MP-QPSI) protocol to output the desired set intersection. In 2024, Chi et al. \cite{17} constructed a QPSI-CA and QPSU-CA protocol for arbitrary tripartite using Bell states and performed circuit simulations to verify the feasibility of the protocols.

Furthermore, with the continuous advancement of QPSI and QPSU, many new combinations and approaches have been proposed. For example, in 2023, Liu et al. \cite{18} proposed a novel QPSI-CA protocol based on a quantum homomorphic encryption (QHE) scheme with Toffoli gates. This protocol was the first to apply quantum homomorphic encryption to the problem of private set intersection. By utilizing the characteristics of homomorphic encryption, it ensures the security of users' private data while addressing the problem of set intersection computation. Then, Mohanty  et al. \cite{19} proposed the first threshold QPSI scheme based on single-particle states and phase encoding. In their protocol, only when the cardinality of intersections reaches or exceeds a predetermined threshold, will the set intersection be revealed. In 2024, Huang et al. \cite{20} designed a  QPSI protocol for multi-party scenarios using single photons and rotational operations, which realizes the intersection of private data sets among multiple users via circular-type transmission. Despite these breakthrough, their protocols rely on $T$ gate encoding or arbitrary phase encoding, which typically demands more complex control and precise phase adjustments \cite{21,22}. This adds significant challenges to the practical implementation of the protocol.
On the other hand, quantum homomorphic encryption offers a unique advantage by enabling secure computations directly on encrypted data without revealing sensitive information. Despite its potential, only Ref.\cite{18} has explored its application in privacy-preserving set computations. However, it relies on homomorphic encryption with Toffoli gates, which poses significant challenges in practical implementation due to its high quantum resource requirements and interaction overhead.
This gap motivates us to design a more efficient and practical approach to privacy-preserving set intersection and union computation using QHE.

In this work, we propose a quantum private intersection and union cardinality protocol based on QHE, which can be divided into four phases. Firstly, a semi-honest third party (TP) and the private set users establish key pairs using the quantum secret sharing protocol, which generates the encryption and decryption keys required for the quantum homomorphic encryption process. Secondly, each private set user transforms their respective privacy data into a sequence of quantum states for subsequent homomorphic encryption. Thirdly, users encrypt the quantum state sequences and send them to TP, where homomorphic evaluation is performed on the sequences by TP. Finally, by applying the decryption keys, the private set comparison result is retrieved and disclosed. The main contributions of this paper are as follows:
\begin{itemize}

\item Based on quantum homomorphic encryption, we propose a quantum approach to address the private set intersection and union cardinality problems. 

\item Leveraging the properties of homomorphic encryption, the proposed protocol enables private set comparisons on encrypted data without prior decryption. Thus, it effectively mitigates the risk of original information leakage.

\item The protocol only requires the execution of simple Clifford gates (i.e., $X$ gate and CNOT gate), avoiding the use of $T$ gate or arbitrary phase encoding, making it feasible with current technological capabilities. The relevant quantum circuits are provided, and circuit simulations are conducted for verification.
\end{itemize}

\section{Preliminaries}

This part provides an overview of the quantum resources used in this paper.

\subsection{Basic of quantum homomorphic encryption}

Quantum homomorphic encryption (QHE) enables operations to be performed on quantum-encrypted data, ensuring that the data remains secure throughout the computation process. The key advantage of QHE is its ability to perform complex quantum computations on encrypted data, which is essential for maintaining privacy in quantum computing environments. Generally, the QHE involves the following four processes\cite{23,24}:
\begin{itemize}
\item  \textbf{Key Generation.} \textit{QHE.KeyGen}: $1^\kappa$ $\rightarrow$($pk,sk,\rho_{evk}$). This process takes the unary representation of the security parameter as input and outputs the classical keys $pk$, $sk$ and a quantum evaluation key $\rho_{evk}$.

\item \textbf{Encryption.}  \textit{QHE.Enc$_{pk}$}: $D (\mathcal{M})\rightarrow D(\mathcal{C})$. By using key $pk$ to transform the message space $\mathcal{M}$ into the cipherspace $\mathcal{C}$.

\item \textbf{Evaluation.} \textit{QHE.Eval$^{QC}_{\rho_{evk}}$}:  $D(\mathcal{C})\rightarrow D(\mathcal{C}^{\prime})$. Based on evaluation key $\rho_{evk}$, 
a quantum evaluation circuit $QC$ is applied to the ciphertext $\mathcal{C}$, and then it produces a new quantum ciphertext state $\mathcal{C}^{\prime}$.

\item \textbf{Decryption.} \textit{QHE.Dec$_{sk}$}: $D(\mathcal{C^\prime})\rightarrow \rho$. Using the private key $sk$, the ciphertext $\mathcal{C}^{\prime}$ is decrypted to obtain the plaintext state $\rho$, where $\rho$ is the result of applying the quantum evaluation circuit to the initial plaintext $D(\mathcal{M})$.
\end{itemize}

\subsection{Homomorphic evaluation requirements for CNOT gate}
This paper primarily relies on the homomorphic evaluation of the CNOT gate, so here we focus on the homomorphic evaluation requirements of CNOT. The matrix form of CNOT, $X$ and $Z$ gates can be expressed as
\begin{equation}
CNOT=
\begin{bmatrix}
1  & 0  & 0 & 0      \\
0  & 1  & 0 & 0      \\
0  & 0  & 0 & 1      \\
0  & 0  & 1 & 0      
\end{bmatrix}, 
X=\begin{bmatrix}
0  & 1      \\
1  & 0 
\end{bmatrix}, 
Z=\begin{bmatrix}
1  & 0      \\
0  & -1 
\end{bmatrix}. 
\end{equation}
These fundamental quantum gates will be used in the following part. 

In the QHE protocol,  the user's quantum information will be encrypted through Pauli operation $X^{a}Z^{b}$, where $a$ and $b$ belong to the key $pk$ and $a, b \in\{0,1\}^n$. After performing the homomorphic evaluation, the decryption key $sk=(\zeta ,\eta)$, ($\zeta, \eta \in\{0,1\}^n$) needs to be updated to achieve the final decryption and obtain the desired result. The key update rule of CNOT gate is shown as follows \cite{25}.
\begin{equation}
\begin{aligned}
(X^{\zeta_{w_i}}, Z^{\eta_{w_i}})= (X^{a_{w_i}},  Z^{b_{w_i}\oplus b_{w_{i+1}}}),\\
(X^{\zeta_{w_{i+1}}}, Z^{\eta_{w_{i+1}}})= (X^{a_{w_i}\oplus a_{w_{i+1}}},Z^{b_{w_{i+1}}}),
\end{aligned}
\end{equation}
where $w_i$ and $ w_{i+1}$ denote the control and target wires. For example,  assuming that there are two quantum states $|\phi_k\rangle$ and $|\psi_l\rangle$, and they are evaluated for CNOT after performing the encryption operation $X^{a_k}Z^{b_k}$ and $X^{a_l}Z^{b_l}$ separately, the result can be expressed as
\begin{equation}
\begin{aligned}
CNOT&(X^{a_k}Z^{b_k} |\phi_k\rangle\otimes X^{a_l}Z^{b_l}|\psi_l\rangle )\\
&\quad=(X^{a_k}Z^{b_k\oplus b_l}\otimes X^{a_k\oplus a_l}Z^{b_l}) |\phi_k\rangle|\phi_k\oplus \psi_l\rangle.
\end{aligned}
\end{equation}


\section{The proposed protocol}

Assuming there are two users, Alice and Bob, who have privacy sets $S_A=\{x_1,x_2,\dots,x_n \} \subseteq  \mathbb{Z}_q$ and $S_B=\{y_1,y_2,\dots,y_n \} \subseteq  \mathbb{Z}_q$, respectively. They want to determine the set intersection and union cardinality with the help of a semi-honest thrid party (TP).  TP is assumed to follow the prescribed steps of the protocol, but she is curious about the user's private data \cite{26,27}. She may take attacks to access the private data of the target user. The proposed protocol consists of four stages and leverages the properties of quantum homomorphic encryption to perform private set comparison. The detailed protocol process is described below (also see Fig. 1).
\begin{figure}[!h]
\centering\includegraphics[width=5.6in]{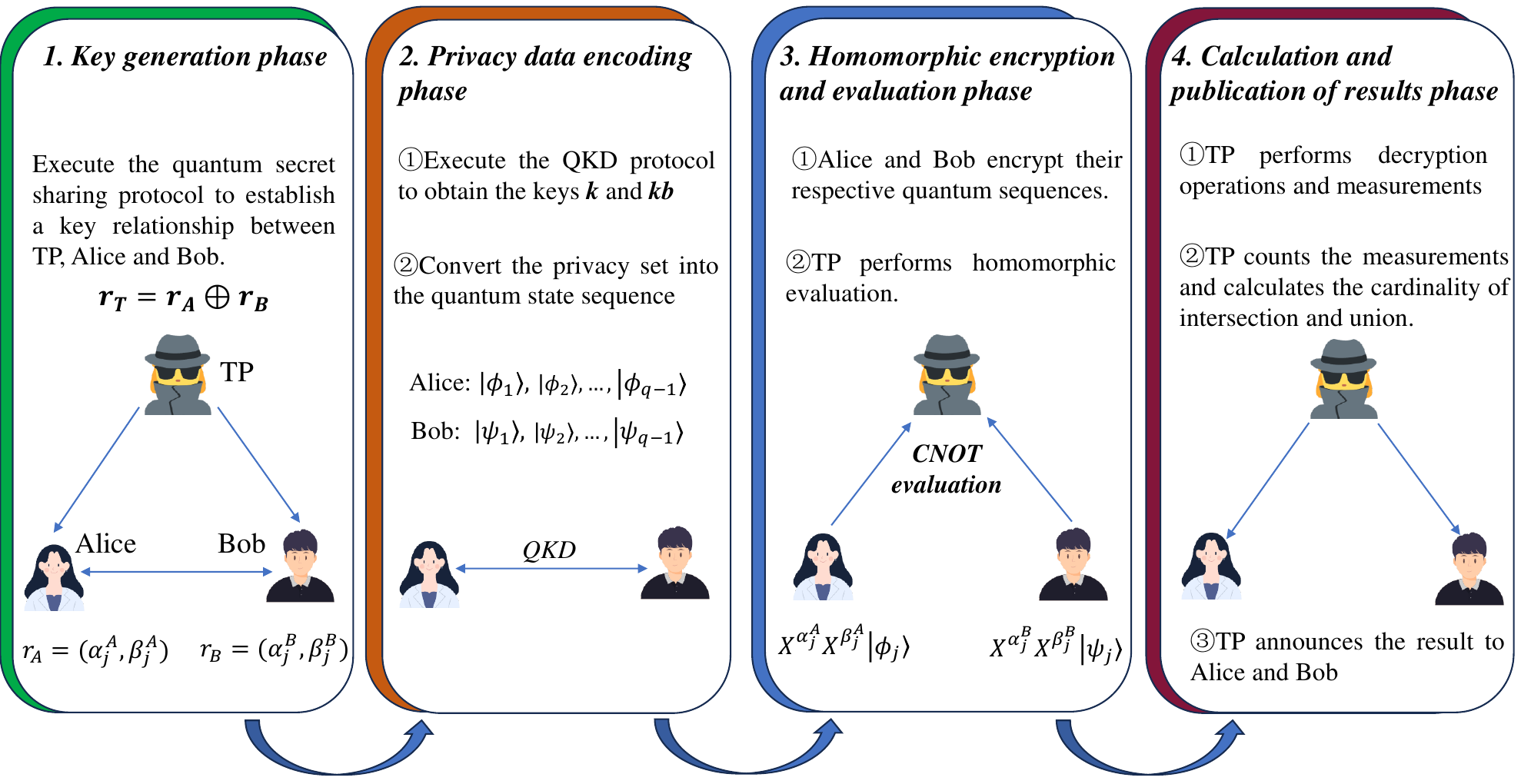}
\caption{Flowchart of the proposed protocol.}
\label{fig:1}       
\end{figure}

\subsection{Key generation phase} TP needs to execute a quantum secret sharing (QSS) protocol with Alice and Bob to establish the key $pk$.  The detailed process is as follows. 

 \textbf{Step 1:} TP first prepares $4q+\delta$ (where $\delta$ is a security parameter) entangled states in the following form \cite{28}:
\begin{equation}
|\Psi\rangle=\frac{1}{\sqrt 2}\left(|0\rangle\frac{|00\rangle+|11\rangle}{\sqrt 2}+ |1\rangle\frac{|01\rangle+|10\rangle}{\sqrt 2} \right).
\end{equation}  
Then, she keeps the first qubit of $|\Psi\rangle$ in her hands and sends the second and third qubits to Alice and Bob, respectively. It should be noted that decoy photons $\{|0\rangle, |1\rangle, |+\rangle, |-\rangle\}$ need to be randomly inserted into the quantum sequence sent by TP to Alice and Bob to ensure the security of transmission.

 \textbf{Step 2:} After receiving the particles, Alice and Bob first conduct an eavesdropping test with TP. Specifically, TP announces the location and preparation basis of the decoy photons, and then Alice and Bob select a suitable measurement basis for measurement and discuss the correctness with TP. If the error rate exceeds the threshold, the protocol terminates, otherwise proceed to the next step.

 \textbf{Step 3:} After discarding the decoy photons, Alice and Bob randomly perform  Z-basis (i.e.,$\{|0\rangle,|1\rangle\}$) or X-basis (i.e., $\{|+\rangle,|-\rangle\}$) measurement on the particles in their hands. For Bell states $\frac{|00\rangle+|11\rangle}{\sqrt 2}$ and $\frac{|01\rangle+|10\rangle}{\sqrt 2}$, performing measurements in the Z-basis and X-basis yields different results, satisfying the following properties: (1) For Bell state $\frac{|00\rangle+|11\rangle}{\sqrt 2}$, the measurement results of both particles are identical when measured in either the Z-basis or the X-basis; (2) For Bell state $\frac{|01\rangle+|10\rangle}{\sqrt 2}$, the measurement results of the two particles are opposite when measured in the Z-basis, but identical when measured in the X-basis. Through these properties, Alice and Bob can detect whether the quantum state $|\Psi\rangle$ prepared by TP complies with the protocol requirements.

\begin{figure}[!h]
\centering\includegraphics[width=3in]{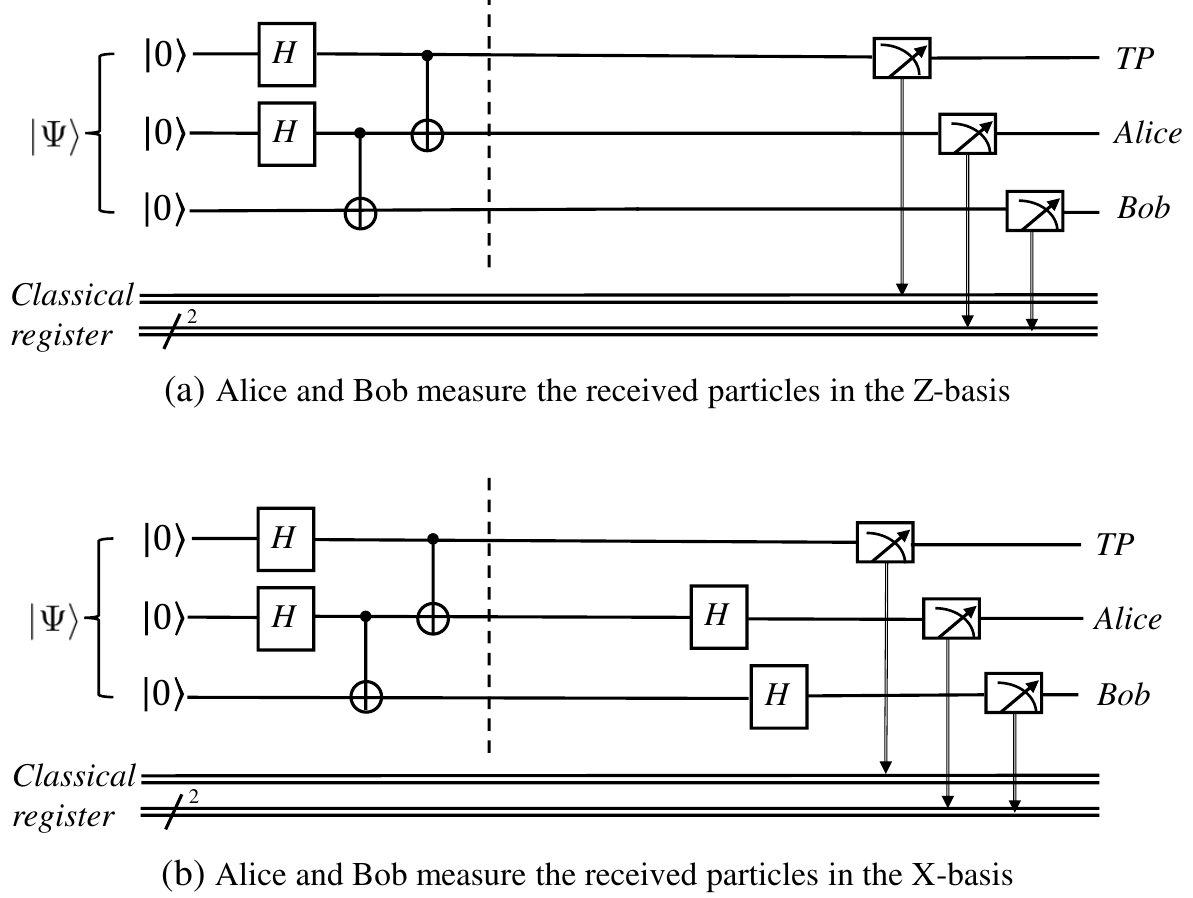}
\caption{ Circuit diagram for performing different measurement basis on the quantum state $|\Psi\rangle$.  The first part of the each circuit, before the dashed line, represents the preparation of state $|\Psi\rangle$, while the second part demonstrates the measurement operations under different bases. In Fig. 2(a), Alice, Bob, and TP each perform Z-basis measurements on their respective particles. In Fig. 2(b), Alice and Bob perform X-basis measurements, while TP performs a Z-basis measurement.}      
\end{figure}

 \textbf{Step 4:}  Alice and Bob randomly select a subset of qubits and instruct TP to publish the measurement results of the qubits kept in her hands. Alice and Bob then discuss the measurements based on their own results as well as those published by TP. Here, they focus only on the case of performing the same measurement basis. Fig. 2 shows the quantum circuits corresponding to the different measurement bases used by Alice and Bob, with further details provided in Table 1. In the cases 1, 2, 3 and 4,  if the error rate exceeds an acceptable threshold, the protocol is terminated.
\begin{table}[htbp]
    \centering
    \caption{Measurement results for different operations}
    \begin{tabular}{cccccc}
        \toprule  
        Case & Alice's operation & Bob's operation & TP's results & Alice's results & Bob's results \\
        \midrule 
        1 & Z-basis & Z-basis &   $|0\rangle$ &  $|0\rangle$ ($|1\rangle$) &   $|0\rangle$ ($|1\rangle$)\\
          \noalign{\smallskip} 
        2 & Z-basis & Z-basis &   $|1\rangle$ &   $|0\rangle$ ($|1\rangle$) &   $|1\rangle$ ($|0\rangle$)\\
          \noalign{\smallskip} 
        3 & X-basis & X-basis &   $|0\rangle$ &   $|+\rangle$ ($|-\rangle$) &   $|+\rangle$ ($|-\rangle$)\\
          \noalign{\smallskip} 
        4 & X-basis & X-basis &   $|1\rangle$ &   $|+\rangle$ ($|-\rangle$) &   $|+\rangle$ ($|-\rangle$) \\
        \bottomrule
    \end{tabular}
\end{table}

\textbf{Step 5:} After security check, Alice and Bob announce which qubits they have measured in the Z-basis. For the state $|\Psi\rangle$, if Alice and Bob perform Z-basis measurements on the second and third particles, respectively, while TP also performs Z-basis measurement on the first particle, they can establish the following key relationships based on their individual measurement results:
\begin{equation}
r_T=r_A\oplus r_B.
\end{equation}  
Note that $r_T, r_A, r_B$ represent the measurement results of TP, Alice and Bob, respectively. 

\textbf{Step 6:} Finally, Alice and Bob use the obtained measurement results $r_A$ and $r_B$ as the key $pk$ for subsequent homomorphic encryption. TP, on the other hand, knows only $r_A\oplus r_B$ and has no access to the separate $r_A$ and $r_B$. This is because TP prepares the corresponding quantum state $|\Psi\rangle$ as required, and she is unable to infer Alice's and Bob's results from the particles retained in her hands.

\subsection{Privacy data encoding phase} At this stage, Alice and Bob's privacy sets ($S_A$ and $S_B$) will be transformed into specific privacy vectors, which are then encoded into corresponding quantum state sequences. The detailed steps are as follows.

\textbf{Step 1:} Alice and Bob first obtain an integer key $k \subseteq \mathbb{Z}_q$ by the QKD protocol \cite{29}. Then, the sets $S_A$ and $S_B$ are transformed into the forms  $S_A*$ and $S_B*$  based on the key $k$: 
\begin{equation}
\begin{aligned}
S_A*=\{ kx_1 \text{mod} \ q,kx_2 \text{mod}\ q,\dots,kx_n \text{mod}\ q   \},\\
S_B*=\{ ky_1 \text{mod} \ q,ky_2 \text{mod}\ q,\dots,ky_n \text{mod}\ q   \}.
\end{aligned}
\end{equation}

\textbf{Step 2:} Alice and Bob execute the QKD protocol again to obtain a binary key $kb=(k^\prime_0,\dots,k^\prime_j,\dots,k^\prime_{q-1})$. Then they  respectively prepare a sequence of quantum states based on the following rules: if $k^\prime_j=0$, Alice and Bob prepare particles 
\begin{equation}
|\phi_j\rangle=\left\{
\begin{aligned}
& |00\rangle \     & \text{if} \ j \not \in S_A*\\
&  |01\rangle \    &  \text{if} \ j \in S_A*
\end{aligned}
\right., 
|\psi_j\rangle=\left\{
\begin{aligned}
& |11\rangle \     & \text{if} \ j \not \in S_B*\\
&  |01\rangle \    &  \text{if} \ j \in S_B*
\end{aligned}
\right.,
\end{equation}
if $k^\prime_j=1$, Alice and Bob prepare particles 
\begin{equation}
|\phi_j\rangle=\left\{
\begin{aligned}
& |11\rangle     & \text{if} \ j \not \in S_A*\\
&  |10\rangle    &  \text{if} \ j \in S_A*
\end{aligned}
\right., 
|\psi_j\rangle=\left\{
\begin{aligned}
& |00\rangle     & \text{if} \ j \not \in S_B*\\
&  |10\rangle    &  \text{if} \ j \in S_B*
\end{aligned}
\right.,
\end{equation}
where $|\phi_j\rangle$ and $|\psi_j\rangle$ denote respectively the $j$-th qubits prepared by Alice and Bob, and $j=0,1,\dots,q-1$.

\subsection{Homomorphic encryption and evaluation phase} Here, Alice and Bob encrypt their respective quantum sequences using the key $pk$ and send them to TP for homomorphic evaluation of the CNOT gate. It should be noted that, since the encrypted particles are $|0\rangle$ and $|1\rangle$, only the $X$ gate is considered here. The $Z$ gate merely changes the phase of the qubit without altering its computational basis state. The focus of the encryption is on whether the quantum state itself changes, rather than phase alteration.

\begin{figure}[htp]
\centering\includegraphics[width=3.5in]{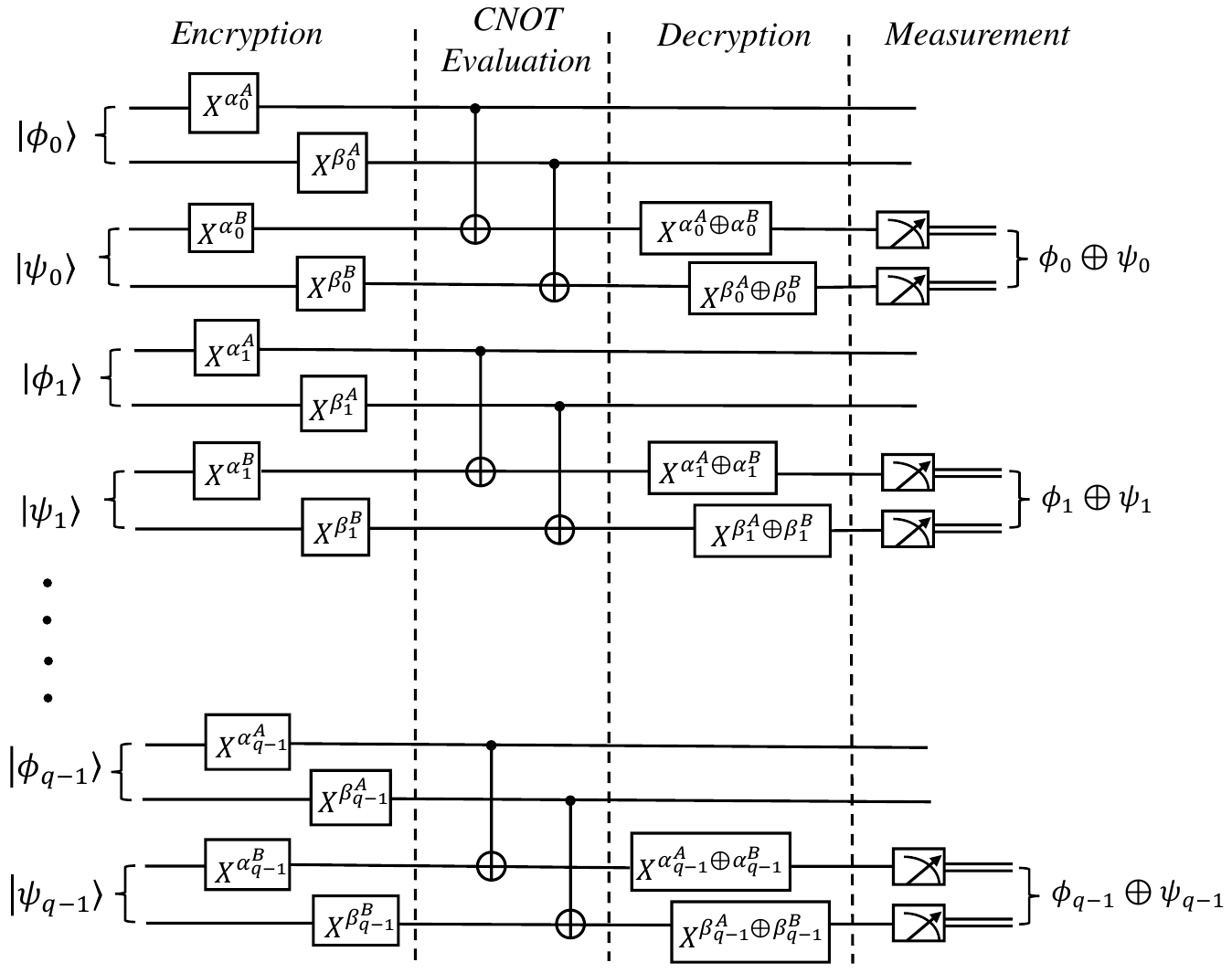}
\caption{Quantum circuit for CNOT homomorphic evaluation. $|\phi_j\rangle$ and $|\psi_j\rangle$ correspond to the particles of Alice and Bob, respectively, with each state containing two particles. This circuit consists of four stages: encryption, CNOT evaluation, decryption, and measurement. In the encryption and decryption phase, the key update rule satisfies the requirements of Eqs. (2) and (3).}
\label{fig:3}       
\end{figure}

\textbf{Step 1:} Alice and Bob use $r_A$ and $r_B$ to encrypt states $|\phi_j\rangle$ and $|\psi_j\rangle$ respectively. The specific method is:
\begin{equation}
|\phi^\prime_j\rangle= X^{\alpha^A_j}X^{\beta^A_{j}}|\phi_j\rangle, \quad |\psi^\prime_j\rangle= X^{\alpha^B_j}X^{\beta^B_{j}}|\psi_j\rangle,
\end{equation}
where ($\alpha^A_j$, $\beta^A_j$) and ($\alpha^B_j$, $\beta^B_j$) represent the  $j$-th group key of $r_A$ and $r_B$, respectively. 

\textbf{Step 2:} Alice and Bob prepare some decoy photons $\{|0\rangle,|1\rangle,|+\rangle, |-\rangle \}$ and then insert them into the sequences $\{|\phi^\prime_0\rangle, \phi^\prime_1\rangle, \dots,\phi^\prime_{q-1}\rangle \}$ and $\{|\psi^\prime_0\rangle, |\psi^\prime_1\rangle, \dots, |\psi^\prime_{q-1}\rangle \}$, respectively, before sending them to TP.

\textbf{Step 3:} Upon receiving the sequences, TP first collaborates with Alice and Bob to perform eavesdropping detection using the decoy photons. If the transmission is deemed secure, TP removes the decoy photons and performs homomorphic evaluation of the CNOT gate on the remaining particles. Specifically, after performing CNOT on the $j$-th $|\phi^\prime_j\rangle$ and $|\psi^\prime_j\rangle$, the state will become 
\begin{equation}
\begin{aligned}
& (CNOT_{2,4}\otimes CNOT_{1,3}) |\phi^\prime_j\rangle |\psi^\prime_j\rangle \\
& = X^{\alpha^A_j} X^{\beta^A_j} |\phi_j\rangle \otimes X^{\alpha^A_j\oplus \alpha^B_j} X^{\beta^A_j\oplus \beta^B_j} |\phi_j\oplus \psi_j\rangle
\end{aligned}.
\end{equation}
 Note that each $|\phi^\prime_j\rangle$ and $|\psi^\prime_j\rangle$ contains two qubits, and $CNOT_{1,3}$ acts on the first qubits of $|\phi^\prime_j\rangle$ and $|\psi^\prime_j\rangle$, while $CNOT_{2,4}$ acts on the second qubits of $|\phi^\prime_j\rangle$ and $|\psi^\prime_j\rangle$. The specific CNOT homomorphic evaluation circuits are given in Fig. 3. For example, consider the qubits $|\phi_0\rangle$ and $|\psi_0\rangle$, which are encrypted using $\{X^{\alpha^A_0}, X^{\beta^A_0}\}$ and $\{X^{\alpha^B_0}, X^{\beta^B_0}\}$, respectively. After performing CNOT evaluation, the corresponding decryption operations are $X^{\alpha^A_0 \oplus \alpha^B_0}$ and $X^{\beta^A_0\oplus\beta^B_0}$. Finally, following the measurement operation, the result is obtained as $\phi_0\oplus\psi_0$.


\subsection{Calculation and publication of results phase} 
After TP completes the homomorphic evaluation, the sequences are decrypted using the updated key to obtain the computation results. TP then counts the different measurements and calculates the cardinality of intersection and union of Alice's and Bob's private sets, and finally publishes the results to Alice and Bob.

\textbf{Step 1:} According to the description of the key generation phase, TP knows the result of $r_A \oplus r_B$. This means that TP can determine the value of 
$\alpha^A_j\oplus \alpha^B_j$ and $\beta^A_j\oplus \beta^B_j$ but cannot know the value of $\alpha^A_j$, $\beta^A_j$, $\alpha^B_j$, $\beta^B_j$ individually. Thus, TP can use $\alpha^A_j\oplus \alpha^B_j$ and $\beta^A_j\oplus \beta^B_j$ as the decryption key $sk$ to decrypt the evaluated quantum state i.e., $| \phi_j\oplus \psi_j\rangle $. 

\textbf{Step 2:} For the $j$-th data item, TP performs the computational basis measurement on the state $| \phi_j\oplus \psi_j\rangle $, and deduces the relationship between Alice and Bob's private data based on the measurement results (see Eqs. (7) and (8) for coding rules). The detailed correspondence is provided in Table 2. 
\begin{table}[htbp]
    \centering
    \caption{Correspondence between measurement results and attribution of privacy data}
    \begin{tabular}{ccc}
        \toprule  
        Measurement results&  Attributed to Alice & Attributed to Bob\\
        \midrule 
        00 &  \checkmark & \checkmark \\
         01 &  \ding{53}  & \checkmark \\
          10 &  \checkmark & \ding{53}  \\
           11 &   \ding{53} &  \ding{53} \\
        \bottomrule
    \end{tabular}
\end{table}

Then, TP uses the variables $h_1$, $h_2$, $h_3$, and $h_4$ to record the number of times that the measurement results are 00, 01, 10, and 11. It is evident that the intersection cardinality of Alice and Bob's private sets is $h_1$, while the cardinality of their union is $h_1+h_2+h_3$.  Finally, TP informs Alice and Bob of the result.

\section{Correctness and Security}

\subsection{Correctness analysis}

\begin{figure}[!ht]
  \centering
  \begin{subfigure}[b]{0.45\linewidth}
    \includegraphics[width=\linewidth,height=3.3cm]{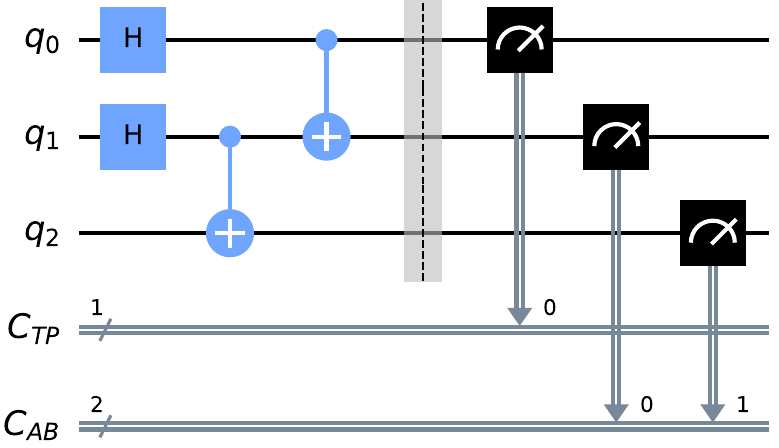}
    \caption{Alice and Bob perform the $Z$-basis measurements}
    \label{fig:sub1}
  \end{subfigure}
  \hspace{0.05\linewidth}
  \begin{subfigure}[b]{0.4\linewidth}
    \includegraphics[width=\linewidth,height=3.6cm]{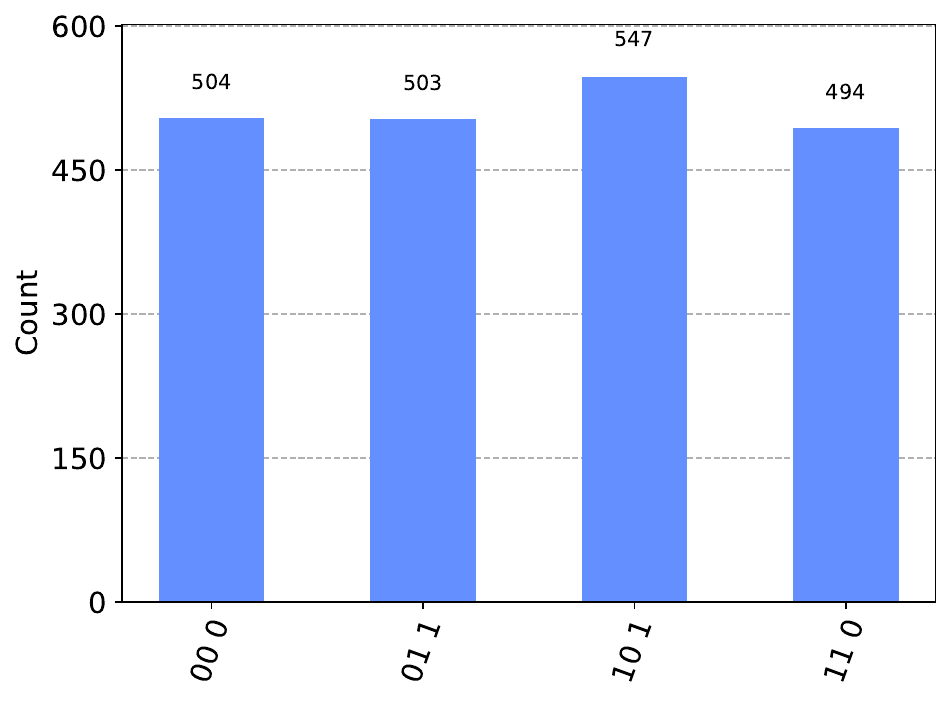}
    \caption{The simulation results of (a).}
    \label{fig:sub2}
  \end{subfigure}
  \hspace{0.05\linewidth}
    \begin{subfigure}[b]{0.45\linewidth}
    \includegraphics[width=\linewidth,height=3.3cm]{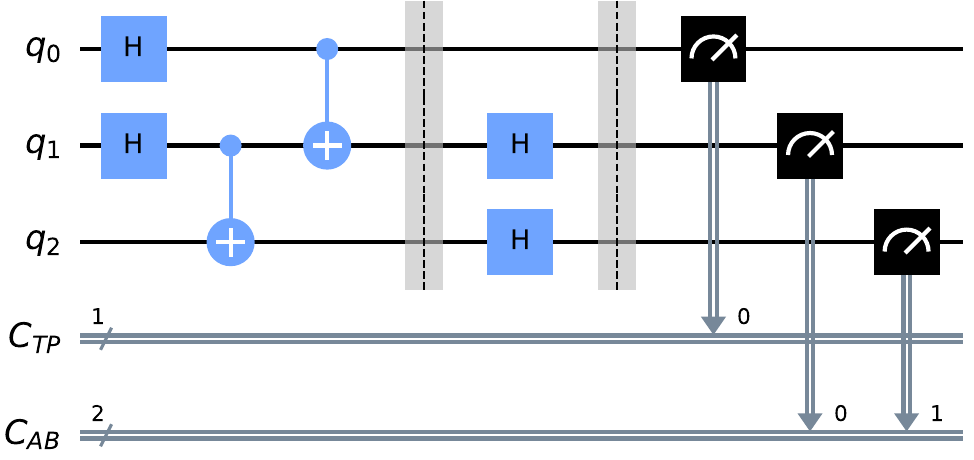}
    \caption{Alice and Bob perform the $X$-basis measurements}
    \label{fig:sub3}
  \end{subfigure}
  \hspace{0.05\linewidth}
  \begin{subfigure}[b]{0.4\linewidth}
    \includegraphics[width=\linewidth,height=3.6cm]{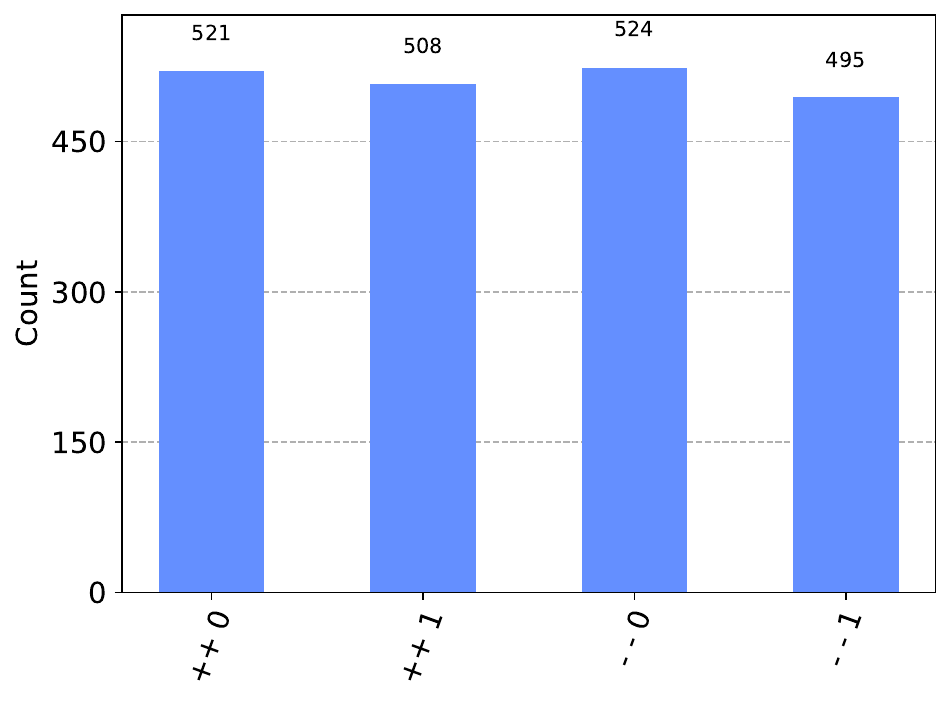}
    \caption{The simulation results of (c).}
    \label{fig:sub4}
  \end{subfigure}
  \caption{Circuits and simulation results of key generation phase. The classic registers $C_{TP}$ and $C_{AB}$ are used to record the measurement results of TP and Alice, Bob, respectively. The quantum registers $q_0$, $q_1$ and $q_2$ represent the particles in TP, Alice and Bob's hands respectively. In the simulation results (b) and (d), the X-axis represents the measurement outcomes, while the Y-axis denotes the frequency of the corresponding outcomes. The total number of simulation runs is 2048.} 
  \label{fig:main}
\end{figure}

To illustrate the correctness of the protocol output, let's consider a specific example. Suppose $S_A=\{1,2,3\}\subseteq  \mathbb{Z}_5$ and $S_B=\{1,2,4\}\subseteq  \mathbb{Z}_5$. The following is an analysis based on the specific protocol stages.

First, during the key generation phase, Alice and Bob will establish a key relationship with TP. Here, we perform a circuit simulation of this phase to verify the correctness of this phase. The specific circuit simulation diagram and results are shown in Fig. 4. 
From Figs. 4(a) and 4(b), it can be observed that when both Alice and Bob choose to perform Z-basis measurements, the measurement results of Alice, Bob, and TP will satisfy the conditions described in Eq. (5) and Table 1. For example, the measurement result in the second column of Fig. 4(b) is $01\  1$, which indicates that TP's measurement result is 1, while Alice and Bob's measurement results are 1 and 0, respectively. This satisfies the condition $1 = 0 \oplus 1$. Additionally, if Alice and Bob choose to perform X-basis measurements, their measurement results will be identical, as clearly shown in Fig. 4(d). For example, in the first column of measurements, both Alice's and Bob's obtained results are $+$. It is also consistent with the steps outlined in the protocol. Consequently, the simulation results align with the design requirements of the key generation phase. After key generation phase, suppose Alice and Bob separately obtain the secret keys:
\begin{equation}
\begin{aligned}
 &(\alpha^A_j,\beta^A_j)= \{(1,1),(0,1),(1,0),(0,1),(1,1)\}, \\
 &(\alpha^B_j,\beta^B_j)=\{(0,1),(1,1),(1,1),(1,0),(0,1)\}, \\
\end{aligned}
\end{equation}
 while TP only knows their XOR result, i.e.,  $(\alpha^A_j\oplus\alpha^B_j,\beta^A_j\oplus\beta^B_j)=\{(1,0),(1,0),(0,1),(1,1),(1,0)\}$. Here, $j=0,1,\dots,4$, correspond to the five results in the key sequence from left to right.

Then in the second phase, suppose Alice and Bob use the key $ \{k = 2\} \subseteq  \mathbb{Z}_5$ to transform $S_A$ and $S_B$ into $S_A*=\{2, 4, 1\}$ and $S_B*=\{2,4 ,3 \}$.  See Eq. (6) for conversion method. Further assume that Alice and Bob establish a binary key $kb = \{k^\prime_0=0, k^\prime_1=1,k^\prime_2= 1,k^\prime_3=1,k^\prime_4= 0 \}$. Based on Eqs. (7-8) and key $kb$, Alice and Bob can respectively generate quantum sequences
\begin{equation}
\begin{aligned}
 &|\phi_0\rangle = |00\rangle, \quad |\psi_0\rangle = |11\rangle, \\
 &|\phi_1\rangle = |10\rangle, \quad |\psi_1\rangle = |00\rangle, \\
 &|\phi_2\rangle = |10\rangle, \quad |\psi_2\rangle = |10\rangle, \\
 &|\phi_3\rangle = |11\rangle, \quad |\psi_3\rangle = |10\rangle, \\
 &|\phi_4\rangle = |01\rangle, \quad |\psi_4\rangle = |01\rangle. 
\end{aligned}
\end{equation}

Subsequently, Alice and Bob encrypt the quantum sequence using the keys obtained in the first stage, as shown in Eq. (9). Then, TP performs the CNOT evaluation, which leads to the result of Eq. (10). Substituting the above assumed specific parameters, we can get the result after homomorphic evaluation
\begin{equation}
\begin{aligned}
X^{1} X^{1} |\phi_0\rangle \otimes X^{1} X^{0}  |\phi_0\oplus \psi_0\rangle, \\
X^{0} X^{1} |\phi_1\rangle \otimes X^{1} X^{0}  |\phi_1\oplus \psi_1\rangle, \\
X^{1} X^{0} |\phi_2\rangle \otimes X^{0} X^{1}  |\phi_2\oplus \psi_2\rangle, \\
X^{0} X^{1} |\phi_3\rangle \otimes X^{1} X^{1}  |\phi_3\oplus \psi_3\rangle, \\
X^{1} X^{1} |\phi_4\rangle \otimes X^{1} X^{0}  |\phi_4\oplus \psi_4\rangle. 
\end{aligned}
\end{equation}
By using  $(\alpha^A_j\oplus\alpha^B_j,\beta^A_j\oplus\beta^B_j)$, TP can decrypt the latter part of the Eq. (13) and subsequently derive the final quantum state $| \phi_j\oplus \psi_j\rangle $. Note that in this example, $j=0,1,\dots,4$.  According to the specific assumptions in the second stage, it is easy to get the result:  $ | \phi_0\oplus \psi_0\rangle =|11\rangle$, $| \phi_1\oplus \psi_1\rangle =|10\rangle$, $| \phi_2\oplus \psi_2\rangle =|00\rangle$, $| \phi_3\oplus \psi_3\rangle =|01\rangle$, and $| \phi_4\oplus \psi_4\rangle =|00\rangle$. To further validate the correctness of the protocol, we conducted a circuit simulation for the cases where $j = 2$, and $j = 4$. The specific circuits
and corresponding results are depicted in Fig. 5.

\begin{figure*}[!t]
\centering
\subfloat[]{\includegraphics[width=3.7in,height=1.7in]{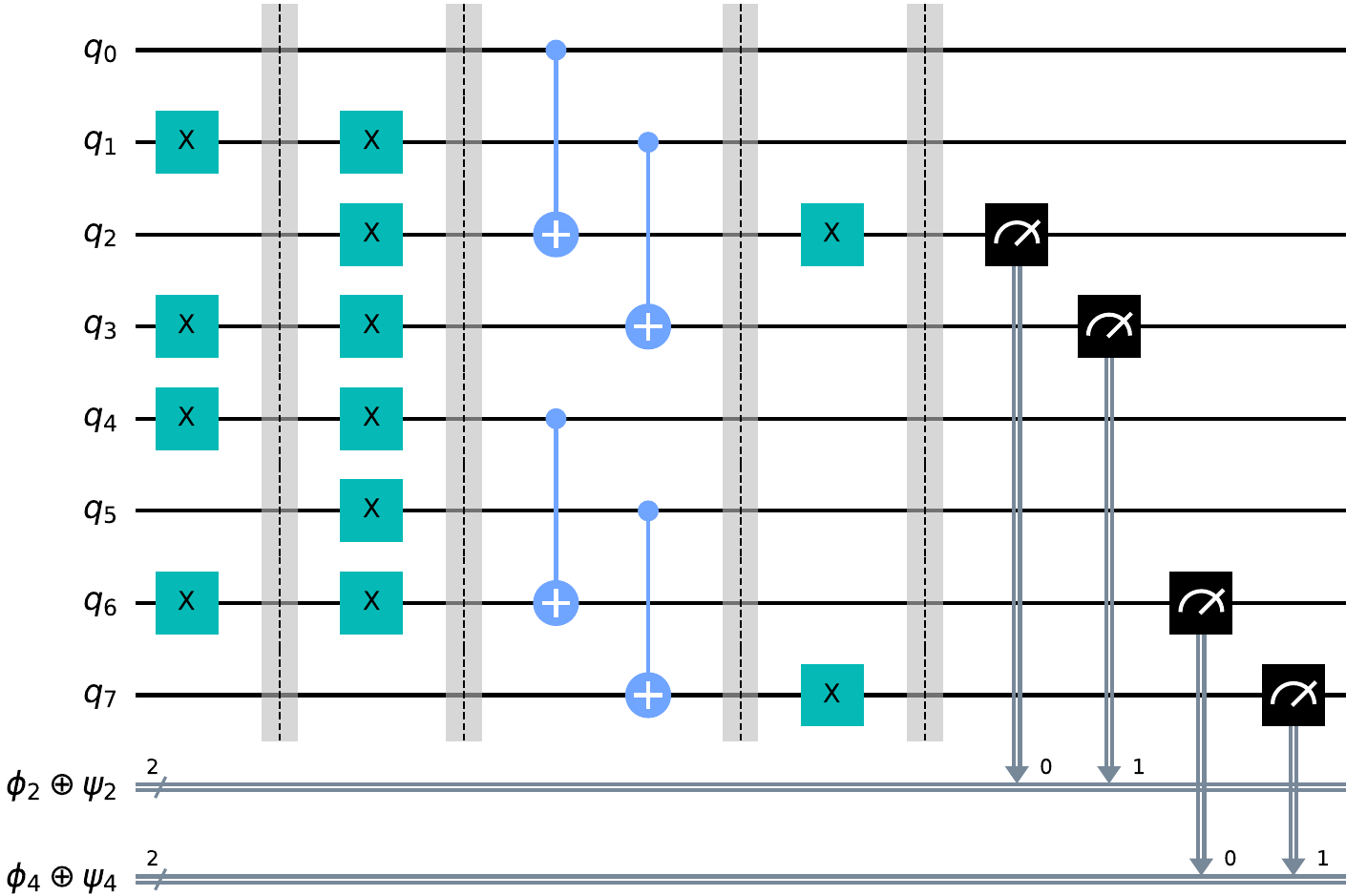}%
\label{fig_first_case}}
\hfil
\subfloat[]{\includegraphics[width=2.3in]{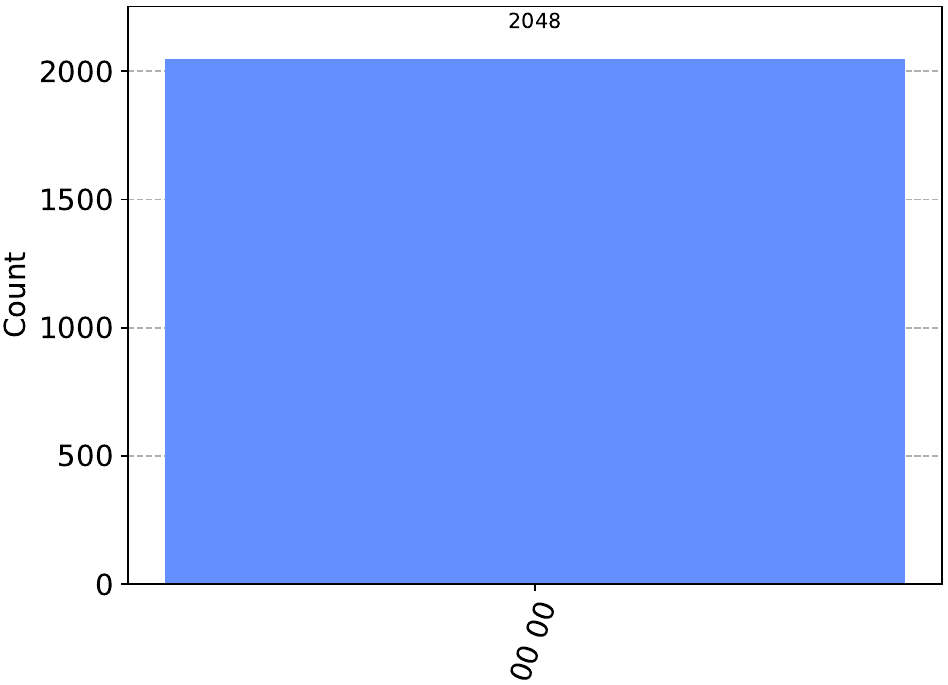}%
\label{fig_second_case}}
\caption{Quantum circuit simulation diagram for the cases of $j=2$ and $j=4$. The circuit includes preparation, homomorphic encryption, CNOT evaluation, decryption, and measurement for quantum states $|\phi_2\rangle=|10\rangle$, $|\psi_2\rangle=|10\rangle$, $|\phi_4\rangle=|01\rangle$, and $|\psi_4\rangle=|01\rangle$. Quantum registers $q_0$ and $q_1$ are used to represent state $|\phi_2\rangle$, $q_2$ and $q_3$ for $|\psi_2\rangle$, $q_4$ and $q_5$ for $|\phi_4\rangle$, and $q_6$ and $q_7$ for $|\psi_4\rangle$.} The simulation result demonstrates that the XOR result of $|\phi_2\rangle$  and $|\psi_2\rangle$ ($|\phi_4\rangle$  and $|\psi_4\rangle$) is 00, which aligns with the protocol's  requirements.
\end{figure*}


After that, TP counts the occurrences of the measurement results $|00\rangle$, $|01\rangle$, $|10\rangle$ and $|11\rangle$, and records them as $h_1$, $h_2$, $h_3$, and $h_4$, respectively. Combining this specific example, we can conclude that $h_1=2$, $h_2=1$, $h_3=1$ and $h_4=1$.  Thus, the cardinality intersection of Alice's and Bob's private sets is $h_1=2$, while the cardinality of their union is $h_1+h_2+h_3=4$. The result obtained is consistent with directly calculating the intersection cardinality of $S_A=\{1,2,3\}$ and $S_B=\{1,2,4\}$, confirming that the output of our protocol is correct.

\subsection{Security analysis}
In this part, we evaluate the security of the proposed protocol, which needs to fulfill the following requirements: (1) The private sets of Alice and Bob must remain confidential and inaccessible to any eavesdropper; (2) TP can only access the cardinality of Alice and Bob's private sets and the inclusion relationships of their union and intersection, but no other information is revealed. The detailed security analysis is as follows.

\begin{theorem}
In the key generation phase, the proposed quantum secret sharing protocol can securely establish the key relationship between TP with Alice and Bob.
\end{theorem} 
\emph{Proof:} In this phase,  TP firstly prepares a sequence consisting of state $|\Psi\rangle=\frac{1}{\sqrt 2}\left(|0\rangle\frac{|00\rangle+|11\rangle}{\sqrt 2}+ |1\rangle\frac{|01\rangle+|10\rangle}{\sqrt 2} \right)$ and uses the decoy photon technique \cite{30} to send the sequence composed of the 2nd and 3rd particles of $|\Psi\rangle$ to Alice and Bob. Due to the insertion of decoy photons, this process can resist typical attacks from external eavesdroppers, including individual and collective attacks. The following takes the entangle-measure attack as an example to illustrate. 

Suppose the attacker Eve uses the unitary operation $U_f$ to perform the entangle-measure attack on the target particle $|\psi\rangle \in \{|+\rangle, |-\rangle,|0\rangle,|1\rangle \}$ with her own auxiliary particle $|0\rangle_E$. The unitary operation $U_f$ is defined as \cite{20} 
\begin{equation}
\begin{aligned}
U_f|x\rangle|y\rangle\rightarrow|x\rangle |y\oplus f(x)\rangle.
\end{aligned}
\end{equation}
Since $U_f$ is unitary, it's easily obtained that $U_f U^\dag_f=I$. If $|\psi\rangle \in\{ |0\rangle, |1\rangle\}$, after $U_f$ operation, we can get
\begin{equation}
U_f|\psi\rangle|0\rangle_E=\left\{
\begin{aligned}
& |0\rangle |f(0)\rangle_E \     & \text{if} \  |\psi\rangle= |0\rangle\\
& |1\rangle |f(1)\rangle_E \    &  \text{if} \  |\psi\rangle= |1\rangle
\end{aligned}
\right..
\end{equation}
On the other hand, if  $|\psi\rangle \in\{ |+\rangle, |-\rangle\}$, after $U_f$ operation, the following result holds:
\begin{equation}
U_f|\psi\rangle|0\rangle_E=\left\{
\begin{aligned}
& \frac{1}{2}\big[\big(|+\rangle+|-\rangle \big)|f(0)\rangle_E + \big(|+\rangle-|-\rangle \big)|f(1)\rangle_E\big] \     & \text{if} \  |\psi\rangle= |+\rangle\\
& \frac{1}{2}\big[\big(|+\rangle+|-\rangle \big)|f(0)\rangle_E - \big(|+\rangle-|-\rangle \big)|f(1)\rangle_E\big] \    &  \text{if} \  |\psi\rangle= |-\rangle
\end{aligned}
\right..
\end{equation}
According to Eqs. (15) and (16), it can be concluded that if Eve's attack is to avoid introducing errors, i.e., to ensure that the state of the decoy particles remains unchanged, then it must satisfy
\begin{equation}
\begin{aligned}
|f(0)\rangle_E=|f(1)\rangle_E=|f\rangle_E.
\end{aligned}
\end{equation}
Thereby, Eqs. (15) and (16) can be rewritten as:
\begin{equation}
U_f|\psi\rangle|0\rangle_E=\left\{
\begin{aligned}
& |0\rangle |f\rangle_E \     & \text{if} \  |\psi\rangle= |0\rangle\\
& |1\rangle |f\rangle_E \    &  \text{if} \  |\psi\rangle= |1\rangle
\end{aligned}
\right.,
\end{equation}
and
\begin{equation}
U_f|\psi\rangle|0\rangle_E=\left\{
\begin{aligned}
& |+\rangle|f\rangle_E  \     & \text{if} \  |\psi\rangle= |+\rangle\\
& |-\rangle|f\rangle_E \    &  \text{if} \  |\psi\rangle= |-\rangle
\end{aligned}
\right..
\end{equation}
From the above equations, it can be observed that regardless of the state of the target particle, if Eve wishes to avoid introducing errors, her ancillary particle must remain independent of the target particle. As a result, she can only obtain identical outcomes from her ancillary particle. Therefore, the protocol is resilient to eavesdropping attacks by attackers.

In the following, we focus on the honesty of TP and the key establishment between Alice, Bob and TP. After security check, Alice and Bob randomly perform Z-basis or X-basis measurements and select test particles at random. They allow TP to release the Z-basis measurement results of the retained qubits to verify whether the prepared state $|\Psi\rangle$ meets the required specifications. Table 1 shows the correct measurement results for each user under different measurement basis. If TP does not prepare the correct quantum state $|\Psi\rangle$, it will be detected by Alice and Bob. Finally, in the case where both Alice and Bob choose Z-basis measurements, and TP also performs Z-basis measurements, their measurement results will satisfy Eq. (5).

 Therefore, TP, Alice, and Bob can establish a secure key relationship, with TP only knowing the XOR results and unable to obtain the individual keys of the users.

\begin{theorem}
The private sets of Alice and Bob remain confidential, even in the presence of a powerful external adversary or a semi-honest TP.
\end{theorem} 
\emph{Proof:} In the proposed protocol, Alice and Bob convert their private sets into the sequences $S_A*$ and $S_B*$ using the secret key $k$ generated by the QKD protocol. These sequences are then encoded into corresponding quantum states for transmission. 
 Therefore, without access to the corresponding key information, an attacker including TP cannot derive the correct private set.
Furthermore, the protocol employs decoy photon techniques to ensure the security of the transmission. If an external adversary attempts to attack the transmitted particles, they will inevitably introduce errors, which can be detected. As for TP, even though she has access to the quantum state sequences of Alice and Bob and performs the CNOT homomorphic evaluation, she cannot obtain specific private information due to the lack of key information $k$. As a result, TP can only determine the cardinality of the intersection or union between Alice's and Bob's datasets, without gaining any detailed private data.

\begin{theorem}
In the homomorphic encryption and evaluation phase, the privacy set of Alice and Bob is hidden in the maximum mixed state. Eavesdroppers cannot obtain useful information in this stage.
\end{theorem} 
\emph{Proof:} In our protocol, Alice and Bob construct a secure key relationship using quantum secret sharing protocol. They then utilize this shared key for homomorphic encryption. During the homomorphic encryption phase, Alice and Bob encrypt their quantum sequences and send them to TP. Specifically, they will perform $X$-gate encryption on qubits $|0\rangle$ or $|1\rangle$ according to the key generated in the key generation phase. After random $X$-gate encryption, the states of $|\phi_j\rangle$ and $|\psi_j\rangle$ will become the maximum mixed state:
\begin{equation}
\begin{aligned}
\sum_{\alpha^A_j ,\  \beta^A_j\in \{0,1\}}\frac{1}{16}X^{\alpha^A_j}X^{\beta^A_j}|\phi_j\rangle \langle \phi_j| (X^{\alpha^A_j}X^{\beta^A_j})^\dag=\frac{I_4}{4},\\
\sum_{\alpha^B_j , \ \beta^B_j\in \{0,1\}}\frac{1}{16}X^{\alpha^B_j}X^{\beta^B_j}|\psi_j\rangle \langle \psi_j| (X^{\alpha^B_j}X^{\beta^B_j})^\dag=\frac{I_4}{4},
\end{aligned}
\end{equation}
where $|\phi_j\rangle$ and $|\psi_j\rangle$ belong to $\{|00\rangle,|01\rangle,|10\rangle,|11\rangle\}$. Hence, without knowing the encryption key, external eavesdroppers or TP cannot obtain the privacy data of Alice and Bob.

\begin{theorem}
Alice and Bob can only learn the intersection and union cardinality of their datasets; no additional information can be accessed.
\end{theorem} 
\emph{Proof:} Without loss of generality, let's assume Alice intends to obtain more information about Bob's private set beyond their intersection and union cardinality. To achieve this, Alice would need to launch an attack during the communication between TP and Bob in an attempt to steal useful information. However, Alice is unaware of the positions of the decoy photons in the transmission sequence between Bob and TP. Consequently, any attack launched by Alice would be detected by both Bob and TP. Moreover, in the final phase, TP only reveals the cardinality  about the intersection and union. Thus, Alice can only know the cardinality from the information disclosed by TP and cannot gain any additional details about Bob's private set. Therefore, Alice cannot obtain any information beyond the intersection union cardinality with Bob's set.

\section{Extension to Multiparty Scenarios}

In this section, the extension to multiparty scenarios is carried out based on the two-party protocol given above.  Suppose there exist $m$ users $A_1, A_2,\dots, A_m$, who have privacy data
\begin{equation}
\begin{aligned}
&S_{A_1}=\{c^{A_1}_1,c^{A_1}_2,\dots,c^{A_1}_n \} \subseteq  \mathbb{Z}_q,\\
&S_{A_2}=\{c^{A_2}_1,c^{A_2}_2,\dots,c^{A_2}_n \} \subseteq  \mathbb{Z}_q,\\
& \quad \vdots \\
&S_{A_m}=\{c^{A_m}_1,c^{A_m}_2,\dots,c^{A_m}_n \} \subseteq  \mathbb{Z}_q,
\end{aligned}
\end{equation}
respectively. They want to determine the set intersection and union cardinality under TP's help. The specific steps of the proposed multiparty protocol are as follows.  

\begin{figure}[!h]
\centering\includegraphics[width=5in]{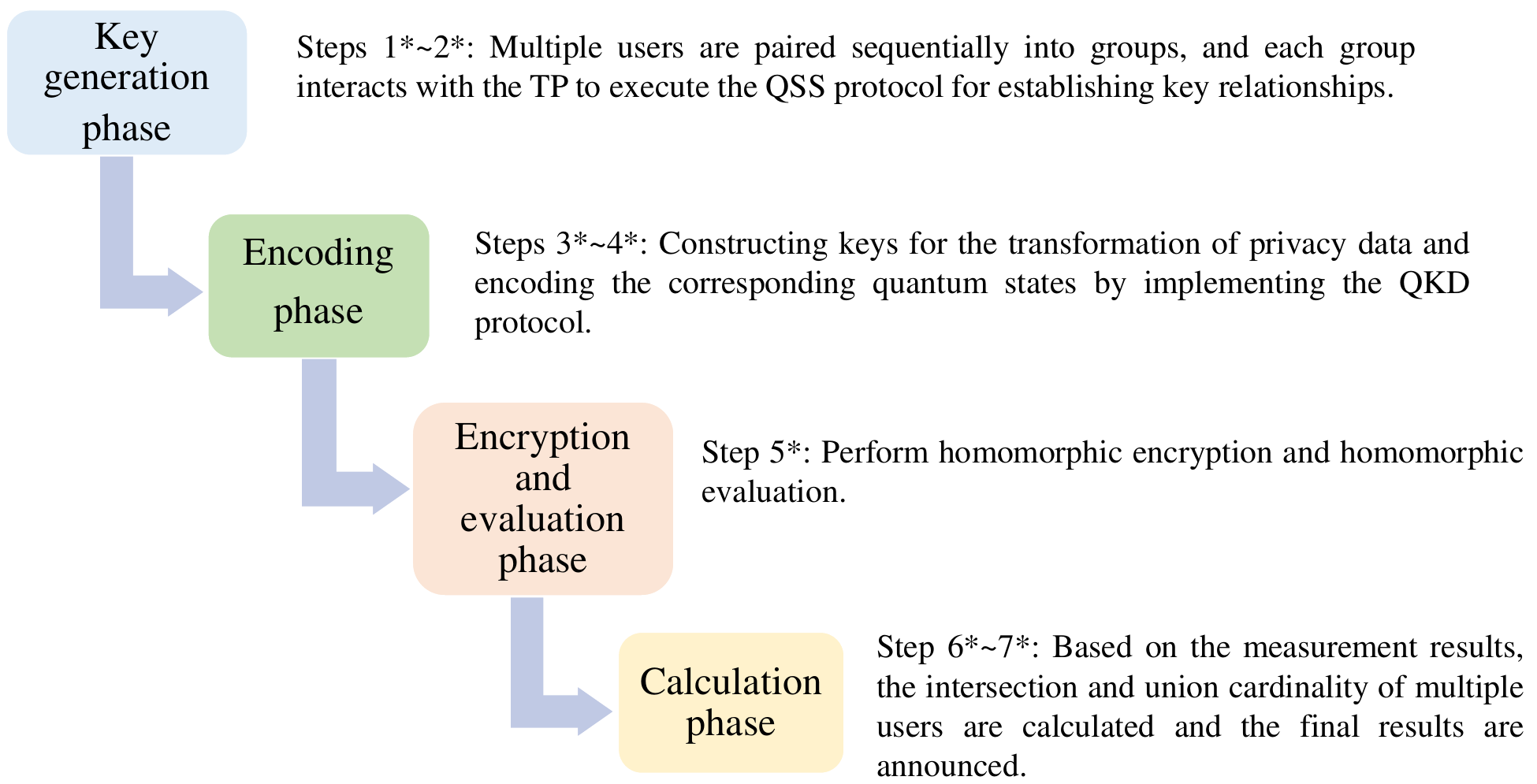}
\caption{The four basic stages of the multi-party protocol and their corresponding steps. }
\label{fig:6}       
\end{figure}

\subsection{Steps of the multi-party protocol}

Similar to the two-party protocol, this protocol consists of four stages: key generation, encoding, encryption and evaluation, and the final computation stage (also see Fig. 6).

 \textbf{Step 1*:}  All users $A_1,A_2,\dots,A_m$ are paired sequentially to form groups, and each group interacts with TP to execute the secret sharing protocol described in Section 3.1 to establish keys. The grouping is arranged as follows: $A_1$ and $A_2$ form the first group, $A_3$ and $A_4$ form the second group, and so on, with $A_{m-1}$ and $A_{m}$ forming the final group.

 \textbf{Step 2*:} After step 1*, each group of users and TP can establish a key relationship similar to Eq. (5). For example, the key relationship established between the group $A_1$, $A_2$, and TP can be expressed as $r_{T_{12}} = r_{A_1} \oplus r_{A_2}$. $A_1$ and $A_2$ will use $r_{A_1}$ and $r_{A_2}$ as $pk$ for subsequent homomorphic encryption, while TP will use $r_{T_{12}}$ as the subsequent decryption key.  The other groups are similar and will not be repeated here.

 \textbf{Step 3*:} All users $A_1,A_2,\dots,A_m$ execute a multi-party QKD protocol \cite{31} to share an integer key $k \subseteq \mathbb{Z}_q$. Then, the private data of  $A_1,A_2,\dots,A_m$ are transformed into 
\begin{equation}
\begin{aligned}
&S^\prime_{A_1}=\{kc^{A_1}_1 \text{mod}\ q,kc^{A_1}_2 \text{mod}\ q,\dots, kc^{A_1}_n \text{mod}\ q \},\\
&S^\prime_{A_2}=\{kc^{A_2}_1\text{mod}\ q,kc^{A_2}_2 \text{mod}\ q,\dots, kc^{A_2}_n\text{mod}\ q \},\\
& \quad \vdots \\
&S^\prime_{A_m}=\{kc^{A_m}_1\text{mod}\ q, kc^{A_m}_2\text{mod}\ q ,\dots, kc^{A_m}_n\text{mod}\ q \}.
\end{aligned}
\end{equation}

\textbf{Step 4*:} Similar to Section 3.2, each group of users, such as $A_1$ and $A_2$, executes the QKD protocol once to obtain a binary key, $kb_{12}=(k^{12}_0,\dots,k^{12}_j,\dots,k^{12}_{q-1})$. Then based on $kb_{12}$ and their respective privacy data (see Eq. 22), $A_1$ and $A_2$ can prepare sequences of quantum states as in Eqs. (7) and (8).

\textbf{Step 5*:} Each group of users (e.g., $A_1$ and $A_2$) performs homomorphic encryption of their respective quantum state sequences using the key established in step 2*, which is then sent to TP for homomorphic evaluation.  The detailed process of homomorphic encryption and evaluation is similar to that described in Section 3.3.

\textbf{Step 6*:} For the two sets of data transmitted by each group, TP decrypts and measures them to obtain the final results. The measurements for each group and their correspondences can be referred to Table 2. Based on the measurement results, TP analyzes each group to determine the relationships between the intersection and union cardinality of the private data of all $m$ users. Specifically, for the $j$-th ($j=1,2,\dots,q-1$) data item:
\begin{itemize}
\item [(1)] \textit{Intersection Determination:} If the corresponding measurement results from all groups are $00$,  it indicates that the value $j$ is part of the intersection of $A_1,A_2,\dots,A_m$. The occurrence of this scenario is recorded using $h^\prime_1$.
\item [(2)] \textit{Not in Any Set:} If the corresponding measurement results from all groups are $11$, it indicates that the value $j$ does not belong to 
$A_1,A_2,\dots,A_m$.
\item [(3)] \textit{Union Determination:} If at least one group's measurement result is $00$, $01$ or $10$, it indicates that the value $j$ is part of the union of $A_1,A_2,\dots,A_m$. The occurrence of this scenario is recorded using $h^\prime_2$.
\end{itemize}

\textbf{Step 7*:} According to the values of $h^\prime_1$ and $h^\prime_2$, TP announces the set intersection and union cardinality to $A_1,A_2,\dots,A_m$.

\subsection{Correctness of the protocol output}

The correctness and security of the protocol in the two-party scenario have been discussed in Section 4. Here we focus on the correctness of the protocol extended to the multi-party scenario.
The entire multi-party protocol is built upon the two-party protocol. To compute the intersection and union among $A_1,A_2,\dots,A_m$, a total of $\lceil \frac{m}{2} \rceil$ two-party protocols need to be executed.

Let $G_1, G_2,\dots,G_{\lceil \frac{m}{2} \rceil}$ denote the $\lceil \frac{m}{2} \rceil$ groups as follows:
\begin{equation}
\begin{aligned}
G_1=\{A_1,A_2\}, G_2=\{A_3,A_4\},\dots,G_{\lceil \frac{m}{2} \rceil}=\{A_{m-1},A_m\}.
\end{aligned}
\end{equation}
For each group $G_l$, we can compute:
\begin{equation}
I^l=\bigcap_{A\in G_l}A ,\quad U^l=\bigcup_{A \in G_l} A,
\end{equation}
which are the intersection and union of the group $G_l$. It is important to note that all users share the same secret key $k$ to encode their private data. As a result, the encoded private data of all users is confined to the set space $[0,q-1]$. Consequently, for any given index $j$, the global intersection and union relationships can be deduced from the intersection and union results of the individual groups. Specifically, for the $j$-th data item, we can calculate
\begin{equation}
\begin{aligned}
&I^1_j=\{A_1 \cap A_2\}_j ,\ I^2_j=\{A_3 \cap A_4\}_j, \ \dots, \ I^{\lceil \frac{m}{2} \rceil}_j=\{A_{m-1} \cap A_m\}_j, \\
&U^1_j=\{A_1 \cup A_2\}_j ,\ U^2_j=\{A_3 \cup A_4\}_j, \ \dots, \ U^{\lceil \frac{m}{2} \rceil}_j=\{A_{m-1} \cup A_m\}_j.
\end{aligned}
\end{equation}
For $j$, if all $I^1_j$, $I^2_j$, to $I^{\lceil \frac{m}{2} \rceil}_j$ are non-empty, it indicates that $j$ belongs to the intersection of $A_1,A_2,\dots,A_m$, i.e., $j\in \bigcap^{m}_{i=1} A_i$. Regarding the union, if at least one $U^l_j$ is non-empty, it implies that $j$ belongs to the union of $A_1,A_2,\dots,A_m$, i.e., $j\in \bigcup^{m}_{i=1} A_i$.

Let us illustrate with a three-party scenario. Assume there are three users, $A_1$, $A_2$ and $A_3$, whose private datasets are $\{1,2,5\}$,  $\{2,3\}$, and  $\{2,4,5\}$, respectively. Additionally, assume that $A_1$, $A_2$ and $A_3$ share a common integer key $k=3 \subseteq  \mathbb{Z}_7$. According to the protocol setup, their private data will be transformed into the forms as $\{3,6,1\}$, $\{6,2\}$, and  $\{6,5,1\}$. Through the analysis and calculations in the two-party scenarios, the intersection and union results for the private sets of $A_1$ and $A_2$, as well as $A_2$ and $A_3$ are as follows: Only when
$j=6$,  it belongs to both $A_1 \cap A_2$ and $A_2 \cap A_3$. Therefore, the cardinality of the intersection of $A_1$, $A_2$ and $A_3$ is 1. Regarding the union, $j=1,2,3,5,6$ are all included in the union of $A_1$, $A_2$ and $A_3$, resulting in a union size of 5.

The result obtained is consistent with directly calculating the intersection cardinality of $\{1,2,5\}$,  $\{2,3\}$, and  $\{2,4,5\}$, confirming that the output of our protocol is correct.

\section{Discussion}

In this section, we discuss the performance of our protocol and compare it with related protocols. It is worth noting that in order to secure users' private data, it is often necessary to execute protocols such as quantum key distribution to construct keys. Such key establishment protocols have also been employed in previous related works\cite{14,15,16,17,18,19,20} to safeguard privacy data.

However, due to variations in the technical approaches employed, we exclude the quantum resources consumed for key establishment from our analysis and comparison for fairness. Instead, we focus on the quantum resources required for computing the intersection and union of private datasets. Furthermore, our analysis is specifically concerned with the quantum-level consumption, excluding the costs of eavesdropping detection.

\subsection{Efficiency and communication complexity}

This part focuses on evaluating and discussing the efficiency and communication complexity of the proposed protocol.

\textit{Qubit efficiency:}  As described in Ref. \cite{17}, the efficiency of qubits can be defined as $\eta=\frac{x}{y+z}$, where $x$ represents the length of the secret information, while $y$ and $z$ denote the total quantum resources utilized and the number of classical bits consumed in the protocol, respectively. We first analyze the protocol's efficiency in the two-party scenario. In our protocol, the private data of each user is encoded into a length of $q$, making $x=q$. Regarding quantum resource consumption, the key generation phase requires TP to prepare $4q$ three-particle entangled states $|\Psi\rangle$. Additionally, in the homomorphic encryption phase, Alice and Bob each consume $2q$ qubits to transmit the encrypted quantum states to TP. Thus, the total quantum resource consumption amounts to $16q$ qubits. Finally, TP needs to tell Alice and Bob the final result, which requires 2 bits of information (i.e., Intersection and Union), so $z=2$. Aggregating all the information, we can conclude that the efficiency of the protocol is $\frac{q}{16q+2}$. Our multi-party protocol is built upon the foundation of the two-party protocol. By applying a similar method of calculation, the final results of qubit efficiency of multi-party protocol can be determined as $\frac{q}{16\lceil \frac{m}{2} \rceil q+2}$, where $m$ denotes the number of users.

\textit{Communication complexity:} In our two-party protocol, $4q$ entangled states $|\Psi\rangle$ are required during the key generation phase to establish a secure key relationship between Alice and Bob. Additionally, during the homomorphic encryption phase, Alice and Bob must each prepare $2q$ qubits for communication with TP. Therefore, the overall communication complexity of the protocol is $\mathcal{O}(q)$. Extending the two-party protocol to a multi-party scenario can be viewed as requiring the execution of $\left\lceil m/2 \right\rceil$two-party protocols. Consequently, the communication complexity of the multi-party protocol can be expressed as $\mathcal{O}(\left\lceil m/2 \right\rceil q)$.


\subsection{Comparison}

This subsection will compare our protocol with related works in terms of quantum resources, protocol functionality, as well as protocol's efficiency and complexity. Detailed comparison results can be found in Table 3.

First, in terms of quantum resources, our protocol primarily utilizes single particles as the information carriers. By executing simple Pauli operations (e.g., $X$ gate) and performing CNOT homomorphic evaluations, it effectively computes the intersection and union of private sets. Compared to prior works (e.g., \cite{14,15,16,17,18,19,20}), our protocol demonstrates advantages in quantum resource usage. Specifically, our protocol requires only basic Pauli and CNOT operations, avoiding the need for complex quantum state manipulations, such as Toffoli gates or quantum phase rotations. Under current technological conditions, executing intricate quantum phase operations still poses significant challenges. For instance, Ref.\cite{18} relies on homomorphic evaluation of Toffoli gates, which involves complex quantum phase gates such as $T$-gates. Similarly, Refs.\cite{19} and \cite{20} rely on complex phase encoding of quantum states. In contrast, our protocol operates solely with Pauli gates and homomorphic CNOT evaluations, making it more practical and easier to implement under existing quantum technologies.

Second, in terms of protocol functionality, our protocol can realize privacy intersection and union cardinality computation, whereas the Refs.\cite{15,16,18,19,20} can only realize privacy intersection computation. In addition, by leveraging the properties of homomorphic encryption, our protocol allows privacy comparisons directly on encrypted data, eliminating the risk of exposing the original data and thereby enhancing the security of the private information. Currently, only our protocol and Ref.\cite{18} employ homomorphic encryption to perform the computation of privacy set data directly on quantum states. In other protocols, measurements are required to convert quantum states into classical information before any comparison can be made. This direct operation on quantum states has the following advantages: it may reduce the number of intermediate steps, avoid the loss of information caused by measurement, and eliminate potential security vulnerabilities.

Third, in terms of efficiency and communication complexity, despite utilizing homomorphic encryption, the efficiency and complexity of our protocol are still comparable. Specifically, in the two-party setting, the efficiency of our protocol is comparable to that of Refs.\cite{14,15,19}. In the multi-party setting, our protocol offers certain advantages compared to Refs.\cite{16,20}. Regarding communication complexity,  our protocol is  also  at a similar level to existing similar protocols. It is worth noting, however, that while our protocol is less efficient than Ref.\cite{18}, the latter relies on Toffoli gates, which, in practice, require user interaction to complete decryption. This additional interaction significantly increases the communication complexity and reduces the overall efficiency of the protocol.

Overall, our protocol is feasible under current technology, as it does not rely on complex quantum states or operations. Furthermore, it enables direct comparison of private set data on encrypted quantum states, enhanced privacy protection.
\begin{table}[htp]
\centering 
\caption{Comparisons between related protocols }
\label{tab:1} 
\begin{threeparttable}      
\begin{tabular}{cccccc}
\Xhline{1.0pt}
\noalign{\smallskip}
 Reference &   Quantum     &   Quantum       &  Applicable    &      Qubit    & Communication\\
           &    resources  &   technologies  &   scenarios    &     Efficiency    & complexity\\
  \hline
  \noalign{\smallskip}  
Ref.\cite{14}  &  GHZ        &  Pauli        &  Three-party  &  $\frac{q}{18q+2}$   &    $\mathcal{O}(q)$\\
               & states      &  operation    &          &                       &     \\
   \noalign{\smallskip}                
Ref.\cite{15}  & Single  &  Pauli     &  Two-party      & $\frac{q}{2q\text{log}q+3q+1}$     &    $\mathcal{O}(q\text{log}q)$\\
               & photons               &  operation &                 &         &     \\
  \noalign{\smallskip}
Ref.\cite{16}  & Single  &  Pauli     &  Multi-party      & $\frac{q}{(m+1)q+1}$     &    $\mathcal{O}(mq)$\\
               & photons &  operation &                 &         &     \\
  \noalign{\smallskip}
Ref.\cite{17}  & Bell    &  Qubit     &  Three-party      & $\frac{q}{15q+1}$     &    $\mathcal{O}(q)$\\
               & states  &  operation &                 &         &     \\
  \noalign{\smallskip}
Ref.\cite{18}  & Single    &   QHE,     &  Two-party      & $\frac{q}{6q+1}$     &    $\mathcal{O}(q)$\\
               & photons              & Toffoli-gate            &                 &         &     \\
  \noalign{\smallskip}
Ref.\cite{19}  & Single  &Phase       &  Two-party      & $\frac{q}{3rq+1}$     &    $\mathcal{O}(rq)$\\
               & photons &encoding        &                 &         &     \\
  \noalign{\smallskip}
Ref.\cite{20}  & Single  & Phase      &  Multi-party      & $\frac{q}{\lambda mq+1}$     &    $\mathcal{O}(\lambda mq)$\\
               & photons & encoding        &                 &         &     \\
  \noalign{\smallskip}
 Our          & Single  &     QHE,     &  Two-party      & $\frac{q}{16q+2}$     &    $\mathcal{O}(q)$\\
 protocol     & photons               &     CNOT-gate        &                 &         &     \\
   \noalign{\smallskip}
Extension of           & Single &     QHE,     &  Multi-party        & $\frac{q}{16\lceil \frac{m}{2} \rceil q+2}$   &   $\mathcal{O}(\left\lceil m/2 \right\rceil q)$\\
 our protocol     &   photons              &     CNOT-gate        &                 &         &     \\
   \noalign{\smallskip}
\Xhline{1.0pt}
\end{tabular}
\begin{tablenotes}
\footnotesize
\item[] QHE: Quantum Homomorphic Encryption, $r$ and $\lambda$: Security parameters, $m$: Number of users, $q$: Size of the private vectors 
\end{tablenotes}
\end{threeparttable}
\end{table}

\section{ Conclusion}
In this work, we present a quantum scheme for private intersection and union cardinality based on quantum homomorphic encryption, accompanied by the corresponding quantum circuits and simulations. Then we analyze the correctness, security, and performance of the proposed protocol, which is further extended to multi-party scenarios. The results demonstrate that it meets the requirements of private intersection and union cardinality protocols and maintains efficiency and communication complexity comparable to the related works. By leveraging the advantages of quantum homomorphic encryption, our approach allows direct operations on the encrypted quantum data without the need for prior decryption, effectively preventing the leakage of original information. Compared to previous protocols, our protocol requires only simple Pauli and CNOT operations for quantum homomorphic encryption, making it feasible under current technological conditions and capable of offering enhanced privacy protection.

Currently, due to the limitations of homomorphic encryption for CNOT gates, our protocol extends to multi-party scenarios primarily by repeatedly executing the two-party protocol to establish the intersection and union relationships among multiple users’ private sets.  In the future research, we will focus on developing advanced methods for achieving simultaneous comparison of multiple users' private sets under the framework of quantum homomorphic encryption. 

\section*{Acknowledgments} 
This work was supported in part by the National Natural Science Foundation of China, No. 61871347.


\end{document}